\documentclass[11pt]{article}

\usepackage{epsfig,latexsym,amsmath}

\newcommand{\myparagraph}[1]{{\smallskip\noindent{\bf #1}}}
\newcommand{\etal}{{\em et al} }

\newcommand{\barx}{{\bar x}}

\newcommand{\calA}{{\cal A}}

\newcommand{\calJ}{{\cal J}}

\newcommand{\braced}[1]{{ \left\{ #1 \right\} }}

\newcommand{\malymax}{{\mbox{\tiny max}}}

\newcommand{\dlast}{{d_\malymax}}

\newcommand{\assign}{{\;\leftarrow\;}}
\newcommand{\addjob}{\oplus }

\newcommand{\mymin}{\mbox{\rm\tiny min}}

\newcommand{\bif}{ \mbox{\bf if} }
\newcommand{\bdo}{ \mbox{\bf do} }
\newcommand{\bthen}{ \mbox{\bf then} }
\newcommand{\bfor}{ \mbox{\bf for} }

\newcommand{\belse}{ \mbox{\bf else} }

\newtheorem{theorem}{Theorem}

\newtheorem{lemma}{Lemma}

\newenvironment{proof}{{\it Proof:\/}}{$\Box$\vskip 0.1in}
        {$\spadesuit$\smallskip}

\newenvironment{bigeqn*}{\large\begin{eqnarray*}}{\end{eqnarray*}}

\begin{document}

\title{A Note on Scheduling Equal-Length Jobs\\
    to Maximize Throughput}

\author{
Marek Chrobak%
\thanks{Department of Computer Science,
    University of California,
    Riverside, CA 92521.
    \{marek,wojtek\}@cs.ucr.edu.
    Supported by NSF grants CCR-9988360, CCR-0208856,
        and NSF/CNRS grant INT-0340752.}
\and
Christoph D\"urr%
\thanks{Laboratoire de Recherche en Informatique,
    Universit\'e Paris-Sud, 91405 Orsay, France. durr@lri.fr.
    Supported by the EU 5th framework programs QAIP
    IST-1999-11234, the NSF/CNRS grant 17171
    and the CNRS/STIC 01N80/0502 and 01N80/0607 grants.}
\and Wojciech Jawor\footnotemark[1]
\and {\L}ukasz Kowalik
\thanks{Instytut Informatyki,
        Uniwersytet Warszawski, Banacha 2, 02--097,
        Warszawa, Poland. \{kowalik,kuros\}@mimuw.edu.pl.
        Supported by KBN grant 4T11C04425.}
\and Maciej Kurowski\footnotemark[3]
}


\renewcommand{\today}{}

\maketitle

\begin{abstract}
  We study the problem of scheduling equal-length jobs with release
  times and deadlines, where the objective is to maximize the number
  of completed jobs. Preemptions are not allowed. In Graham's
  notation, the problem is described as $1|r_j;p_j=p|\sum U_j$. We
  give the following results: (1) We show that the often cited
  algorithm by Carlier from 1981 is not correct. (2) We give an
  algorithm for this problem with running time $O(n^5)$.
\end{abstract}


\section{Introduction}

We study the following scheduling problem: We are given $n$ jobs
numbered $1,2,\dots,n$. For each job $j$, a release time $r_j$ and a
deadline $d_j$ are given. All jobs have the same processing time
$p$. (We assume that all numbers are positive integers.)  We want to
find a non-preemptive schedule of the given set of jobs that maximizes
the throughput, where the \emph{throughput} is defined as the number
of completed jobs. (Equivalently, we can minimize the number of late
jobs.) In Graham's notation, the problem is described as
$1|r_j;p_j=p|\sum U_j$.

The feasibility version of this problem, where we ask whether
\emph{all} jobs can meet their deadlines, has been studied thoroughly.
Polynomial time algorithms for this version were first found,
independently, by Simons~\cite{simons78} and Carlier~\cite{carlier81}.
A faster algorithm, with running time $O(n \log n)$, was subsequently
given by Garey \etal~\cite{garey81}.  Interestingly, all three
algorithms use quite different techniques.

The elegant feasibility algorithm of Carlier \cite{carlier81} is based
on a dynamic programming approach that processes jobs from left to
right (in order of release times).  For each time $t$, it constructs a
partial schedule with jobs that complete execution at or before time
$t$.  Carlier also considers a certain \emph{dominance} relation on
partial schedules. Intuitively, one partial schedule dominates another
if this other schedule cannot be extended to produce an overall better
schedule, independently of the jobs that are to be released after time
$t$. The schedule computed for each time has the property that it
dominates all other partial schedules up to this time.

In \cite{carlier81}, Carlier also attempted to extend his technique to
the maximization problem, and proposed a polynomial-time
algorithm. This result is now widely cited in the literature. However,
in Section~\ref{sec: carlier's algorithm} we present an example of
an instance on which Carlier's algorithm produces a sub-optimal schedule,
proving that this algorithm is not correct.
We then extend our construction to
show that even the general approach from \cite{carlier81}
does not work. To this end, we show that any left-to-right
dynamic programming algorithm needs to keep track of an exponential
number of partial schedules. (See Section~\ref{sec:no_left_right}
for a more rigorous statement of this claim.)
This result  reveals an interesting feature
of scheduling equal-length jobs, as it shows that the maximization
problem is structurally more difficult than the feasibility problem.

Finally, in Section~\ref{sec: A Maximization Algorithm}, we present an
$O(n^5)$-time algorithm for $1|r_j;p_j=p|\sum U_j$. Our technique is
based on the approach developed by Baptiste \cite{Baptiste}, who gave
an $O(n^7)$-time algorithm for the weighted version of this problem.


\section{Preliminaries}
        \label{sec: preliminaries}

In the rest of the paper $\calJ$ denotes a set of $n$ jobs, $r_j$ and
$d_j$ the integer release time and deadline of a job $j$, and $p$ a
fixed processing time.  Typically, we assume $\calJ =
\braced{1,2,\dots,n}$, although in the next section we also use capital
letters $A,B,\dots$, to denote jobs (possibly with indices.)

We order the jobs according to the deadlines, breaking the ties
arbitrarily, that is, $i<j$ implies $d_i \le d_j$, for all $i,j$.
Without loss of generality, we assume that $d_j \ge r_j + p$ for all
$j$. We can further assume that $\min_i r_i = 0$, and by
$\dlast = d_n$ we denote the latest deadline.

We say that a job $j$ is \emph{executed} in time interval $[t,t']$ if
it is started and completed in $[t, t']$. (Equivalently, the job $j$
is started after or at $t$, but no later than at $t'-p$.)  Thus all the
scheduled jobs must be executed in the interval $[0,\dlast]$.

A \emph{schedule} $S$ is a function that assigns starting times to
some of the jobs in $\calJ$, such that the scheduled jobs are executed
between their release times and deadlines and no two jobs overlap.

Define a schedule $S$ to be \emph{canonical} if it has the following
two properties:
\begin{description}
\item{(c1)} $S$ is \emph{left-shifted}, in the following sense:
        for any job $j \in S$, the starting time of $j$ is
        either $r_j$ or
        the completion time of the job in $S$ that precedes $j$.
\item{(c2)} $S$ is \emph{earliest-deadline}, that is,
        if $i,j\in S$ and $i$ is scheduled before $j$,
        then either $i$ is scheduled before $r_j$ or $i < j$.
\end{description}
It is quite easy to see that each schedule can be converted into a
canonical schedule without changing the set of completed jobs.
For suppose that there are two jobs $i,j\in S$ that violate
condition (c2). This means that $i>j$, but $i$ is scheduled at or
after time $r_j$ and before $j$. We can then swap $i$ with $j$ in
the schedule. We
continue this process until we obtain a schedule that satisfies (c2).
Then, to achieve condition (c1), we modify the schedule from left
to right
by shifting all jobs leftward, either to their release times, or to
the completion time of the previous job, whichever is greater.
Note that this shifting cannot violate property (c2).

For simplicity, we will slightly abuse the terminology above
and treat a schedule simply as a \emph{sequence} of jobs. The
actual schedule corresponding to this sequence is a left-shifted
schedule that executes the jobs in the given order.
Further, we will sometimes treat a schedule $S$ as a
\emph{set} of scheduled jobs, and write $i\in S$, etc.

Denote by $C(S)$ the makespan of a schedule $S$ (that is, the latest
completion time of a job in $S$.) If $m\not\in S$ and $d_m \ge
C(S)+p$, then by $S \addjob m$ we denote the schedule obtained from
$S$ by scheduling $m$ at time $\max\braced{C(S), r_m}$. The schedule
$S \addjob m$ is called an \emph{extension} of $S$.


\section{Carlier's Algorithms}
        \label{sec: carlier's algorithm}

\paragraph{Feasibility algorithm.}
In \cite{carlier81} Carlier presented two algorithms for scheduling
equal-length jobs. The first algorithm was for determining
feasibility, and the second for maximizing throughput.

The feasibility algorithm for each time $x=0,1,\dots,\dlast$ computes a
schedule $S_x$ of jobs to be executed in the interval $[0,x]$.
We call $S_x$ \emph{active} if it contains all jobs with deadlines
at most $C(S_x)$.  The schedules $S_x$ are computed
incrementally, from left to right, as follows. At time $x$, we
consider the set $H = \braced{j\in \calJ: r_j\le x-p} - S_{x-p}$,
namely the jobs that were released at or before $x-p$ and have not
been scheduled in $S_{x-p}$. If $H= \emptyset$, we take $S_x =
S_{x-p}$. Otherwise, pick the earliest-deadline job $m \in H$. If
$S_{x-p}\oplus m$ is active, let $S_x = S_{x-p}\oplus m$,
otherwise $S_x =S_{x-1}$.
After we compute $S_{\dlast}$, if $S_{\dlast} = \calJ$, we are done,
otherwise report that $\calJ$ is not feasible. To achieve polynomial
time, we can modify the algorithm, so that it only considers the time
values $x = r_j + lp$, for some $j\in\calJ$ and $l\in\braced{0,\dots,n}$.

To justify correctness of this algorithm,
Carlier considers a certain \emph{dominance} relation on partial
schedules. Intuitively, one partial schedule dominates another if this
other schedule cannot be extended to produce an overall better
schedule, independently of the jobs that are to be released in the
future. It turns out that, under the assumption that $\calJ$ is
feasible, at each step $x$, schedule $S_x$ dominates all other
schedules in the interval $[0,x]$. In particular, this implies that,
for $x=\dlast$, the resulting schedule $S_\dlast$ will contain all jobs
from $\calJ$. It should be
noted that this algorithm, as well as the next maximization algorithm,
does not take into account the values of the deadlines of the
non-expired jobs, when choosing the next job to schedule, only their ordering.


\paragraph{Maximization algorithm.}
In \cite{carlier81}, Carlier also attempted to extend his technique to
the maximization problem, and proposed the following algorithm:

\begin{center}
\begin{minipage}{14cm}
\begin{tabbing}
\hskip 0.2in\=\hskip 0.3in\=\hskip 0.3in\=\hskip 0.3in\=\hskip 0.3in\=\kill
{\sc Algorithm~1}\\
Initialize $S_x^0 = \emptyset $ for all $x$ and $S_x^k$ undefined for $k\ge1$\\
\bfor $k = 1,2,\dots,n$ \bdo\\
\>\bfor $x = p,p+1,\dots,\dlast$ \bdo\\
\>\> $H\assign\braced{j\in \calJ: r_j+p\le x} - S_{x-p}^{k-1}$\\
\>\> $H'\assign\braced{j\in H:  d_j \ge x}$ \\
\>\> \bif $H' = \emptyset$\\
\>\>\> \bthen $S_x^k \assign S_{x-1}^k$\\
\>\>\> \belse\\
\>\>\>\> $m \assign$  earliest deadline job in  $H'$\\
\>\>\>\> $S_x^k \assign S_{x-p}^{k-1} \addjob m$\\
return $S_\dlast^k$ for the largest $k$ such that $S_\dlast^k$ is defined
\end{tabbing}
\end{minipage}
\end{center}
The algorithm does not specify the start times of the
jobs in the schedule, only their ordering. As explained
in the introduction, the actual schedule is obtained
by left-shifting the jobs in this sequence.

\begin{figure}[ht]
\noindent
\begin{minipage}{2.5in}
\centerline{\epsfig{file=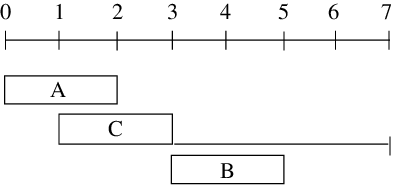,width=5cm}}
\end{minipage}
\hfill
\begin{minipage}{4in}
{\small
\begin{tabular}%
{p{0.4in}|rp{0.195in}p{0.195in}p{0.195in}p{0.195in}p{0.195in}p{0.195in}p{0.195in}p{0.195in}}
$S^k_x$ & $x =$& 0 & 1 & 2 & 3 & 4 & 5 & 6 & 7 \\ \hline
$k\!=\!1$ && - & - & A & C & C & B & C & C \\
$k\!=\!2$ && - & - & - & - & AC & CB & CB & BC \\
$k\!=\!3$ && - & - & - & - & - & - & - & -
\end{tabular}
}
\end{minipage}
\caption{Counter-example for Carlier's maximization algorithm.
The instance is shown on the left, and the schedules
$S^k_x$ are shown in the table on the right.}
\label{fig:counter_ex}
\end{figure}

\myparagraph{A counter-example.}
We now show that the above algorithm is
not correct. Consider the instance $\calJ = \braced{A,B,C}$ given in
Fig.~\ref{fig:counter_ex}. Each job $j$ is represented by a rectangle
of length equal to the processing time (in this case $p=2$), extending
from its release time $r_j$ to $r_j+p$, and by a line segment
extending from $r_j+p$ to $d_j$. For example, $r_C = 1$ and $d_C = 7$.

In the loop for $k=1$ the algorithm constructs the schedules
$(A)$, $(C)$, and $(B)$. Then, for $k=2$, it computes the
schedules $(A,C)$, $(C,B)$, $(B,C)$, and for $k=3$ it finds no
schedules. The optimal schedule is $(A,B,C)$. The algorithm fails
simply because the schedules computed for $k=2$ do not include
$(A,B)$, which is the only schedule of 2 jobs that can be extended
to $(A,B,C)$ by adding a job at the end.


\section{A Better Counter-Example}\label{sec:no_left_right}

The example above still leaves open the possibility that the algorithm
can be corrected with some minor modifications.  We now give another
construction showing that even the general strategy from
\cite{carlier81}, namely a \emph{left-to-right dynamic programming
  approach}, as defined later in this section, will not work. In this
approach we associate with every time $t$ a set ${\bf S}_t$ of
partial schedules, which are idle after $t$. Every set ${\bf S}_t$
is computed by adding more jobs to schedules in sets ${\bf S}_{t'}$
for some earlier times $t'<t$. Further, the decisions of the algorithm
at time $t$ can only depend on the values of the release times and 
deadlines that are smaller or equal $t$ and comparisons involving 
deadlines that are greater than $t$.

\begin{figure}[htb]
\centerline{\epsfig{file=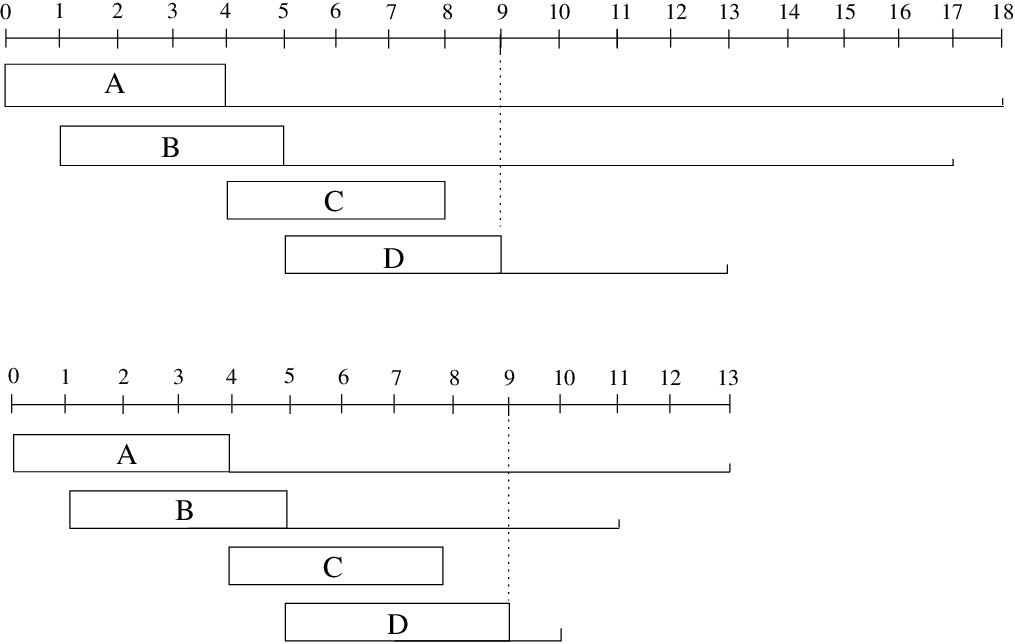,width=9cm}}
\caption{An example of two instances where a left-to-right dynamic
programming algorithm must store two schedules at time $t=9$.}
\label{fig:incomparable}
\end{figure}

To illustrate this concept, consider the instances shown in
Figure~\ref{fig:incomparable}, where $p=4$.
Up to time $9$ the two instances are indistinguishable.
The overall optimal schedules are $(A,C,D,B)$ for the upper instance, and 
$(B,D,A)$ for the lower instance. So a left-to-right dynamic
program must store the partial schedules $(B,D)$ and $(A,C)$ at time $9$.

We now amplify this construction by telescoping $n$ instances of
this kind. As a result, we get an exponential number of partial
schedules in some initial interval $[0,t_0]$ that are
``indistinguishable" from each other by any left-to-right dynamic
programming algorithm. Since the optimal solutions to those
instances in $[0,t_0]$ are different, any such algorithm would
have to keep track of exponentially many partial schedules.

For every $m$-bit string $\barx = x_0x_1\ldots x_{m-1}$
we define an instance $\calJ_\barx$ that consists of $4m$ jobs. The
instance is partitioned into $m$ sub-instances, with sub-instance
$i\in\braced{0,\dots,m-1}$ containing jobs $A_i$, $B_i$, $C_i$, and
$D_i$.  We take $p$, the processing time, to be a sufficiently large
integer (as explained later, any $p\ge 2m+3$ will work). For
$i=0,\dots,m$, let
\begin{eqnarray*}
        u_i \;=\; i(2p+1) \quad\mbox{\rm and}\quad
        v_i \;=\; m(2p+1) + \sum_{j=i}^{m-1} (p+(p+1)x_i).
\end{eqnarray*}
Let $t_0=u_m=v_m$.  The time-scale is divided into intervals
$[u_i,u_{i+1}]$ for the left part and into intervals $[v_{i+1},v_i]$
for the second part.  The release times and deadlines of the jobs are
given in the following table.

\begin{figure}[t]
  \centering \epsfig{file=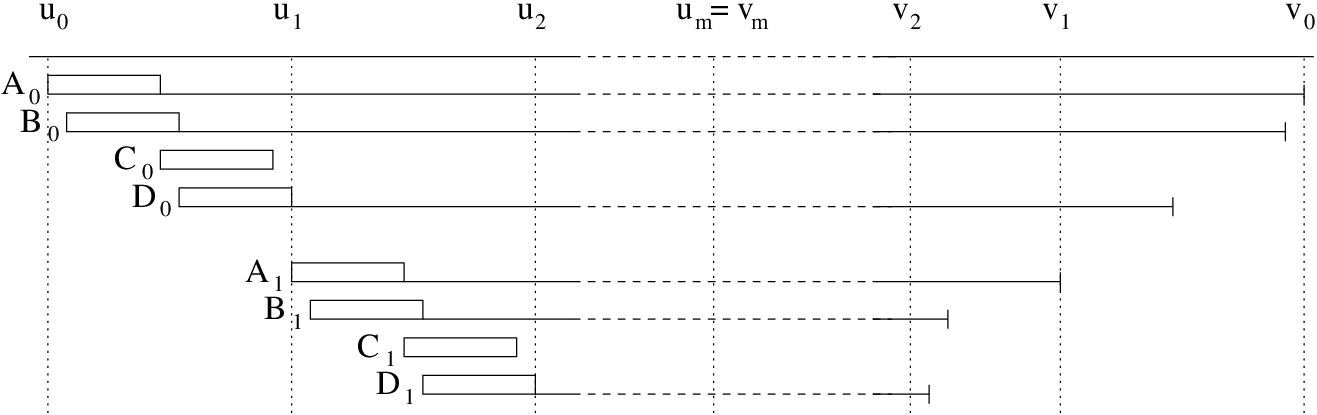,width=6.4in}
  \caption{The instance $\calJ_\barx$ defined by a
                 bit string $\barx$ starting with $10$.}
  \label{fig:blocks}
\end{figure}

\begin{table}[h]
\begin{center}
\begin{tabular}{l|llllll}
job         & $A_i$          &$B_i$       &$C_i$    &$D_i$    \\
\hline
release time& $u_i$          &$u_i+1$     &$u_i+p$  &$u_i+p+1$\\
deadline if $x_i = 0$&
              $v_{i+1}+p$    &$v_{i+1}+2$ &$u_i+2p$ &$v_{i+1}+1$\\
deadline if $x_i = 1$&
              $v_{i+1}+2p+1$ &$v_{i+1}+2p$&$u_i+2p$ &$v_{i+1}+p$
\end{tabular}
\end{center}
\end{table}
\noindent
Figure~\ref{fig:blocks} shows an example of an instance $\calJ_\barx$,
for $\barx = 10...$.


For any given $m$-bit string $\barx = x_0x_1\ldots x_{m-1}$, we define
a schedule $R_\barx$ of $\calJ_\barx$ as follows for all $0\le i\le
m-1$:
\begin{itemize}
\item If $x_i = 0$,  $B_i$ and $D_i$ are scheduled at their
        release times and $A_i$ right before its deadline
        (that is, at time $v_{i+1}$.) $C_i$ is not scheduled.
\item If $x_i=1$, $A_i$ and $C_i$ are scheduled at their release
        times and $D_i, B_i$ right before their deadlines (that is, at
        times $v_{i+1}$ and $v_{i+1}+p$, respectively.)
\end{itemize}
%


\begin{lemma}\label{lem: opt schedule}
  Let $\barx = x_0x_1\ldots x_{m-1}$ be an $m$-bit string. Then
  $R_\barx$ is an optimal schedule for $\calJ_\barx$.  Moreover, any
  optimal schedule $S$ for $\calJ_\barx$ contains the same job
  sequence as $R_\barx$.
\end{lemma}

\begin{proof}
  By routine inspection, $R_\barx$ is a correct schedule. We claim
  that $R_\barx$ is optimal.  Let $\xi = \sum_i x_i$. Then schedule
  $R_\barx$ contains $3m+\xi$ jobs. In this schedule each interval
  $[u_i,u_{i+1}]$ has idle time $1$, and each interval $[v_{i+1},v_i]$
  has idle time $x_i$. So the total idle time is $m+\xi\le 2m$, and
  thus, for $p\ge 2m+3$, $R_\barx$ is an optimal schedule.

  It remains to show that every optimal schedule has the same job
  sequence as $R_\barx$.  For this purpose let $\bar S$ be an
  optimal schedule and let $S$ be a schedule obtained from $\bar S$
  by left-shifting the jobs before $t_0$ and right-shifting the jobs after
  $t_0$.  Formally every job scheduled before
  $t_0$ either starts at its release time or at the completion time of
  another job, and every job scheduled at $t_0$ or later completes
  either at its deadline or at a time where another job is started.
  Clearly the order in which the jobs appear in $\bar S$ and $S$ is
  the same.

  By induction, we show for every $0\le i\le m-1$ that $S$ is
  identical with $R_\barx$ in $[u_i,u_{i-1}]\cup[v_{i+1},v_i]$.  By
  induction hypothesis these intervals can contain no job from
  $A_j,B_j,C_j,D_j$ for $j<i$.  It shows that the only available jobs
  in $[u_i,u_{i-1}]\cup[v_{i+1},v_i]$ are $A_i,B_i,C_i,D_i$.

  When $x_i=0$, the interval $[v_{i+1},v_i]$ must contain the job
  $A_i$, otherwise the idle time would be more than $p-2>2m$, which
  contradicts the optimality of $S$.  Then the interval $[u_i,u_{i+1}]$
  cannot contain $C_i$, since this would imply that the period
  $[u_i,u_i+p]$ is idle.  Therefore this interval must contain the
  jobs $B_i,D_i$, in that order.

  When $x_i=1$, the intervals $[u_i,u_{i+1}]$ and $[v_{i+1},v_i]$ must
  contain all four jobs, at least portions of it, otherwise the idle
  time would be at least $p>2m$.  Job $C_i$ is tight.  Therefore job
  $A_i$ must be scheduled at its release time.  This in turn shows that
  $D_i,B_i$ must be scheduled in the interval $[v_{i+1},v_i]$ in
  that order.
\end{proof}

A left-to-right dynamic algorithm is an algorithm $\calA$ with the
following properties:
\begin{itemize}
\item It processes the jobs from left to right on the time axis,
and for each time $t$ it constructs a collection ${\bf S}_t$ of
partial schedules in the interval $[0,t]$. The schedules in ${\bf
S}_t$ are obtained by extending schedules from ${\bf S}_{t'}$, for
$t' < t$, by appending new jobs at the end. The final schedule is
chosen from ${\bf S}_t$ for $t = \dlast$.
\item At each step $t$, $\calA$ decides which non-expired jobs should
  be added to previous schedules. This decision is based on values
  of the release times and deadlines which are smaller or equal $t$ and on
  pairwise comparisons between deadlines that are greater than $t$.
\end{itemize}

Consider now the behavior of $\calA$ on instances $\calJ_\barx$.
Until $t_0$, all these instances are indistinguishable to $\calA$,
since all jobs expired by time $t_0$ are identical, and the
deadline ordering for the jobs. (For each $i$, the deadline
ordering of $A_i, B_i,D_i$, does not depend on whether $x_i = 0$
or $1$.) Thus at time $t_0$, for each $\barx$, $\calA$ would need
to store a partial schedule that consist of the first $2m$ jobs
from $R_\barx$. So $\calA$ would need to keep track of
exponentially many partial schedules.

\medskip

It should be noted that the maximization algorithm proposed by
Carlier does fall into the framework described above. Although the
algorithm, as shown, makes $n$ left-to-right passes along the time
axis, each schedule $S^k_t$ depends only on schedules
$S^{k'}_{t'}$ for $k' \le k$ and $t' \le t$. So the algorithm
could be equivalently reformulated to compute all these schedules
in just a single pass.

\medskip

One may also ask what happens if some bounded look-ahead, say
$\ell p$, is allowed in the algorithm. Our construction can be
easily modified to show that this will also not work. To see this,
simply add $\ell$ ``tight" jobs to the instances, with release
times $t_0,t_0+p,\dots,t_0+(\ell-1)p$, and deadlines
$t_0+p,\dots,t_0+\ell p$, respectively, and shift the right half
of the instances by $\ell p$. The same result holds, if the
algorithm is granted the ability to verify if the schedules in
${\bf S}_t$ are feasible or not. The details are left to the
reader.


\section{A Maximization Algorithm}
\label{sec: A Maximization Algorithm}

Our algorithm is based on a technique developed by Baptiste
\cite{Baptiste} for solving the weighted version of the problem. We
show how to improve Baptiste's $O(n^7)$ time complexity exploiting the
fact that the jobs have equal weights.
Our algorithm can be thought of as a dual to the one in \cite{Baptiste}.
While Baptiste's algorithm computes maximum weight schedules for
a certain family of sub-instances,
our algorithm considers similar sub-instances, but for given
weights computes \emph{minimum makespans} of schedules with given
weights. In our case, the weight of a schedule is the number of
jobs, and thus is bounded by $n$. This gives us a desired
reduction in running time.


\begin{theorem}
The scheduling problem $1|r_j;p_j=p|\sum U_j$ can be solved in time $O(n^5)$.
\end{theorem}

\begin{proof}
Our algorithm uses dynamic programming and it runs in $n$ phases.
During the $k$-th phase we take under consideration only jobs $1, 2,
\ldots, k$. (Recall that jobs are ordered by their deadlines.)
Let $\Theta' = \braced{r_i + lp : \; i=1, \ldots, n; \;
l=-1,0,\ldots, n}$.  It is easy to see that $|\Theta'| = O(n^2)$.  For
each $\alpha\in \Theta'$ and each $u \in \braced{1 \ldots k}$ we
define:
\begin{description}
\item{$\calJ^k_\alpha$\;=\;} the set of all jobs $j\in\braced{1,2,\dots,k}$
                with $r_j \ge \alpha$.
\item{$B^k_{\alpha, u}$\;=\;}
the minimal value $\beta \in \Theta'$ such that it is possible to
execute exactly $u$ jobs from $\calJ^k_\alpha$ in the interval
$[\alpha+p,\beta]$.
\end{description}
If a schedule in the definition of $B^k_{\alpha, u}$ does not exist,
we assume that $B^k_{\alpha, u} = +\infty$. It is convenient to extend
this notation by setting $B^k_{\alpha, 0} = \alpha +p$ for all $k \in
\braced{0, \ldots, n}$, $\alpha \in \Theta'$, and $B^k_{\alpha, u} =
+\infty$ for $k \in \braced{0, \ldots, n}$, $\alpha \in \Theta'$ and
$u > k$.

Now we show how $B^k_{\alpha, u}$ depends on the values computed
during the previous phase, i.e.  $B^{k-1}_{\alpha', u'}$, for $\alpha'
\in \Theta'$ and $u' \in \braced{0, \ldots, n}$.  Let $k > 0$,
$\alpha$, $u > 0$ be fixed, and let us consider a canonical schedule
$S$ that satisfies the conditions defining $B^k_{\alpha, u}$:
\begin{description}
\item{(s1)} exactly $u$ jobs from $\calJ^k_\alpha$ are executed in $S$,
\item{(s2)} all jobs are executed in the interval $[\alpha+p, \beta]$,
\item{(s3)} $\beta$ is minimal,
\end{description}
where $\beta = B^k_{\alpha, u}$.
If $k\notin S$ then we have $B^k_{\alpha, u} = B^{k-1}_{\alpha,
u}$. So from now on we will assume that $k\in S$, which implies $r_k\geq \alpha$. We will denote the
number of jobs executed in $S$ before and after $k$ by $x$ and $y$,
respectively. The starting time of job $k$ is denoted by
$\gamma$. (See Fig~\ref{fig:div}.)

\begin{figure}[t]
\begin{center}
\epsfig{file=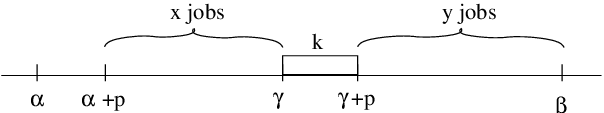,width=8cm}
\end{center}
\caption{Notation}
\label{fig:div}
\end{figure}

Since the jobs are ordered by their deadlines, and $S$
is canonical, we conclude that $S$ must satisfy the following
conditions:
\begin{description}
\item{(s4)} all the jobs executed after job $k$
        are in $\calJ^{k-1}_\gamma$ and are executed
        in the interval  $[\gamma+p,\beta]$,
\item{(s5)} all the jobs executed before job $k$
        are in $\calJ^{k-1}_\alpha$ and are executed
        in the interval $[\alpha+p,\gamma]$,
\item{(s6)} $\gamma = \max( r_k, B^{k-1}_{\alpha, x})$,
\item{(s7)} $\beta = B^{k-1}_{\gamma, y}$.
\end{description}
Notice that, if we knew the value of $x$, we could easily derive
$\gamma$ using property (s6), and then derive $\beta$ using property
(s7).  As we do not know $x$, we have to iterate over all possible
values $x = 0, 1, \ldots, u-1$, and choose the schedule with minimal
$\beta$. Finally, we set $B^k_{\alpha, u}$ to $\min(B^{k-1}_{\alpha,
u}, \beta)$.

The algorithm is summarized in the pseudo-code below:

\begin{center}
\begin{minipage}{14cm}
\begin{tabbing}
\hskip 0.2in\=\hskip 0.3in\=\hskip 0.3in\=\hskip 0.3in\=\hskip 0.3in\=\kill
{\sc Algorithm~2} \\
Let $\Theta'$ be the set of values $r_i + pl$
where $i\in\braced{1,2,\dots,n}$ and  $l \in \braced{-1, 0, \ldots, n}$ \\
\bfor $\alpha \in \Theta'$,
      $k \in \braced{0, \ldots, n}$, and
      $u \in \braced{k+1, \ldots, n}$ \bdo $B^k_{\alpha,u} \assign +\infty$ \\
\bfor $\alpha \in \Theta'$ and
        $k \in \braced{0, \ldots, n}$ \bdo $B^k_{\alpha, 0} \assign \alpha + p$ \\
\bfor $k \assign 1,2,\dots,n$ \bdo \\
\> \bfor $\alpha \in \Theta'$ and
        $u \in \braced{1, \ldots ,k}$ \bdo \\
\>\> $\beta_{\mymin} \assign B^{k-1}_{\alpha, u}$ \\
\>\> \bif $r_k \geq \alpha$ \bthen\\
\>\>\> \bfor $x \assign 0,1,\dots,u-1$ \bdo \\
\>\>\>\> $y \assign u-x-1$ \\
\>\>\>\> $\gamma \assign \max(r_k, B^{k-1}_{\alpha,x})$ \\
\>\>\>\> \bif $\gamma+p\le d_k$ \bthen\\
\>\>\>\>\>    $\beta \assign B^{k-1}_{\gamma,y}$ \\
\>\>\>\>\>    $\beta_{\mymin} \assign \min(\beta_{\mymin}, \beta)$ \\
\>\> $B^k_{\alpha, u} \assign \beta_{\mymin}$ \\
\end{tabbing}
\end{minipage}
\end{center}

After the completion of the algorithm, the number of jobs executed in
the optimal schedule is equal to the maximal value of $u$ such that
$B^n_{-p, u} \neq +\infty$. (Recall that, by our convention,
$\min(\Theta') = -p$.) The optimal schedule itself can also be easily
derived from the values stored in the array $B$ with the use of
standard techniques.

The overall complexity of our algorithm is $O(n^5)$, because each of
$O(n^4)$ values $B^k_{\alpha,u}$ is computed in linear time.
\end{proof}

\bibliographystyle{abbrv}
\bibliography{scheduling}

\end{document}